\documentclass[aps,showpacs,preprint,preprintnumbers,amsmath,amssymb,nofootinbib]{revtex4}
\usepackage{epsfig}
\usepackage{graphicx}
\usepackage{dcolumn}
\usepackage{bm}
\usepackage{feynmp}
\usepackage{slashed}

\def\beq{\begin{equation}}
\def\eeq{\end{equation}}

\newcommand{\bea}{\begin{eqnarray}}
\newcommand{\eea}{\end{eqnarray}}

\def\eeqn{\end{equation}}
\newcommand\iden{\leavevmode\hbox{\small1\normalsize\kern-.33em1}}

\let\jnfont=\rm
\def\NPB#1,{{\jnfont Nucl.\ Phys.\ B }{\bf #1},}
\def\PLB#1,{{\jnfont Phys.\ Lett.\ B }{\bf #1},}
\def\EPJC#1,{{\jnfont Eur.\ Phys.\ Jour.\ C }{\bf #1},}
\def\PRD#1,{{\jnfont Phys.\ Rev.\ D }{\bf #1},}
\def\PRL#1,{{\jnfont Phys.\ Rev.\ Lett.\ }{\bf #1},}
\def\MPLA#1,{{\jnfont Mod.\ Phys.\ Lett.\ A }{\bf #1},}
\def\JPG#1,{{\jnfont J.\ Phys.\ G }{\bf #1},}
\def\CTP#1,{{\jnfont Commun.\ Theor.\ Phys.\ }{\bf #1},}
\def\JHEP#1,{{\jnfont JHEP \ }{\bf #1},}
\def\NPPS#1,{{\jnfont Nucl.\ Phys.\ Proc.\ Suppl.\ }{\bf #1},}

\newcommand{\rpv}{R_p\!\!\!\!\!\! /\,\,}

\begin{document}


\title{LHCb $\bigtriangleup A_{CP}$ of $D$ meson and R-Parity Violation}

\author{{Xue Chang}$^1$, {Ming-Kai Du}$^1$,
{Chun Liu}$^1$, {Jia-Shu Lu}$^1$, {Shuo Yang}$^{2,3}$}
\affiliation{
$^1$State Key Laboratory of Theoretical Physics,
Institute of Theoretical Physics, Chinese Academy of Sciences,
P.O. Box 2735, Beijing 100190, P.R. China\\
$^2$Department of Physics, Dalian University, Dalian 116622, P.R. China\\
$^3$Center for High-Energy Physics, Peking University, Beijing, 100871, P.R. China}
\email{chxue@itp.ac.cn, mkdu@itp.ac.cn, liuc@mail.itp.ac.cn,
lujiashu@itp.ac.cn, yangshuo@dlu.edu.cn}

\begin{abstract}

LHCb collaboration has recently announced a measurement of
the difference of time-integrated CP asymmetries between
$D\rightarrow K^+K^-$ and $D \rightarrow \pi^+\pi^-$. This
result provides the evidence of large direct CP violation
in $D$ meson and reveals some important implications on underlying
new physics. It is shown that the
direct CP violation in $D$ meson can be enhanced
by R-parity violating supersymmetry, while
CP violations in $K$ and $B$ mesons are suppressed by this new
physics, which is in consistence with previous experiments.
Constraints on the model parameters and some consequences
are also discussed.

\end{abstract}



\maketitle
\section{Introduction}
New physics might be discovered first through direct searches
at colliders, or via an indirect way, i.e., be observed
in precision measurements at 'low' energy. The key motivations of
CP violation (CPV) measurements at LHCb are just precision
tests of the Standard Model (SM) and searching for new physics.
The CPV in $D$ meson is highly suppressed in SM,
which hence provides a background-free search for new physics.
Furthermore, the hadron built with charm quark is the only
playground of CPV in $u$-type quark sector because the top
quark decays before it could be hadronized. Hadrons built with
$u$ or $\bar{u}$, such as $\pi ^0$ and $\eta$, are their own
antiparticles, therefore no CPV occurs in these systems.

Recently, LHCb collaboration has announced a
measurement of the difference between CP asymmetries in two
$D$ meson decay channels \cite{exp},
\begin{equation}
\begin{aligned}
   \bigtriangleup A_{CP}^{\mathrm{dir}}&\equiv
   A_{CP}(D^0\to K^+ K^-)-A_{CP}(D^0\to \pi^+ \pi^-)\\
   &=[-0.82\pm0.21(\mathrm{stat}.)\pm0.11(\mathrm{syst})] \%~.~
\end{aligned}
\end{equation}

This measurement make it robust against
systematics and is mainly sensitive to direct CPV.
This result deviates significantly from the prediction of
SM, in which it is at the order of $10^{-4}$
\cite{review,review2,DCPV in charm,Ddecay}.
Although ATLAS and CMS have not found any evidence of new
physics, this large $\bigtriangleup A_{CP}^{\mathrm{dir}}$
at LHCb still can provide a hint of underlying new physics.

In this work, we presented a tentative interpretation of the
enhancement of direct CPV in $D$ meson with R-parity
violating (RPV) supersymmetry, while leaving that
of $K$ and $B$ mesons nearly unaffected, since the SM predictions
of CPV in $K$ and $B$ mesons are consistent with previous
experiments. In Sec. II, we gave a brief estimate of the direct CPV in
SM, through which some essential RPV parameters were obtained. Then in
Sec. III, we listed our conclusion and discussed some relevant implications.

\section{$\rpv$ -SUSY and Direct CP violation in $D$ decay }
Before going to R-parity violating supersymmetry, let us make
a brief review of SM calculation for this CPV
\cite{review,review2,DCPV in charm,Ddecay}. In the SM, CP violations in
$D^0(\bar D^0)\to\pi^+ \pi^-$ and $D^0(\bar D^0)\to K^+ K^-$ decays are significantly
suppressed by CKM parameters, loop effects, and GIM mechanism.
At the quark-gluon level, the $\pi^+ \pi^-$ case is
depicted in Fig. 1, and the $K^+K^-$ case by the same diagrams with the
replacement of $d \rightarrow s$.  CPV in the decays is due to the
interference between the tree amplitude $\mathcal{M}_T^{SM}$ (Fig.1 left)
and the penguin diagram amplitude $\mathcal{M}_P^{SM}$ (Fig.1 right).
It is defined as \cite{CPRPV}
\begin{equation}
\begin{aligned}
\label{ACP}
   A_{CP}^{\mathrm{dir}}\equiv
   \frac{\Gamma-\bar{\Gamma}}{\Gamma+\bar{\Gamma}}
   &\simeq \frac{\sum_{T\leftrightarrow P}(-2)
   \mathrm{Im}( \alpha_T^{*\mathrm{SM}} \alpha_P^\mathrm{SM} )
   \mathrm{Im} (\mathcal{M}_T^{*\mathrm{SM}} \mathcal{M}_P^\mathrm{SM})}
   {|\alpha_T^\mathrm{SM}|^2|\mathcal{M}_T^\mathrm{SM}|^2}~,~
\end{aligned}
\end{equation}
where
\begin{equation}
\begin{aligned}
   &\mathcal{M}^{\mathrm{SM}}(D^0\to\pi^+ \pi^-)
   =\alpha_T^{\mathrm{SM}} \mathcal{M}_T^{\mathrm{SM}}
   +\alpha_P^{\mathrm{SM}} \mathcal{M}_P^{\mathrm{SM}}~, \\
   &\alpha_T^{\mathrm{SM}} (D^0\to\pi^+ \pi^-)=V_{ud}V_{cd}^*~,\\
   &\alpha_P^{\mathrm{SM}} (D^0\to\pi^+ \pi^-)=-V_{ub}V_{cb}^*~ .
   \end{aligned}
\label {eq: parameter1}
\end{equation}

To $\alpha_s$ order, $A_{CP}^{\mathrm{dir}}$ can be simplified as
\begin{equation}
   A_{CP}^{\mathrm{dir}}(\mathrm{SM})\simeq
   \frac{-2\mathrm{Im}(\alpha_P^{\mathrm{SM}})
   \mathrm{Im}(\mathcal{M}_P^{\mathrm{SM}})}
   {\alpha_T^{\mathrm{SM}} \mathcal{M}_T^{\mathrm{SM}}}~.~
\label {eq: direct cp}
\end{equation}

\vspace{1pt}
\begin{figure}[htbp]
\centering
\includegraphics[width=3.5in]{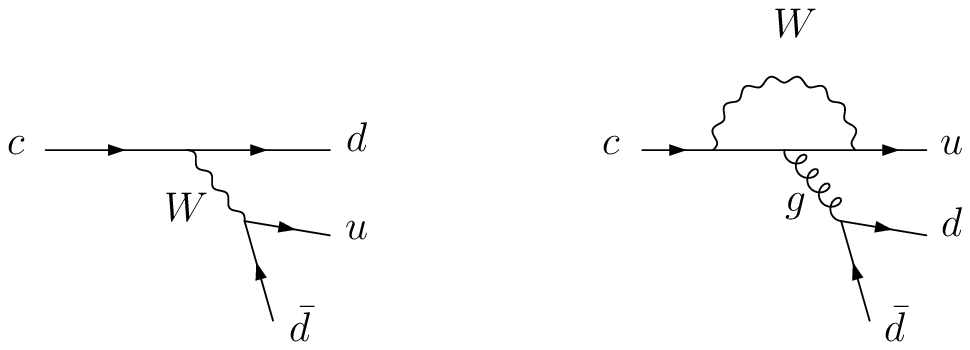}
\caption{$c\rightarrow\bar{d}du$ tree level and penguin diagrams in the SM.}
\end{figure}

The tree level diagram amplitude is
\begin{equation}
\begin{aligned}
  \mathcal{M}_T^{\mathrm{SM}}(D^0\to &\pi^+ \pi^-)\\
  &=i\frac{G_F}{\sqrt{2}}
   \langle\pi^-|\bar{d}\gamma^\mu c|D^0\rangle
   \langle\pi^+|\bar{u}\gamma^\mu \gamma_5 d|0\rangle
   \approx-\frac{G_F}{\sqrt{2}}m_D^2f_+(m_\pi^2) f_\pi~,~
\end{aligned}
\label {eq: tree}
\end{equation}
where the hadronic matrix elements are parameterized as
\begin{equation}
\begin{aligned}
  &\langle \pi^+|\bar{u}\gamma^\mu \gamma_5 d|0\rangle
    =if_\pi p_{\pi^+}^\mu~,~\\
   &\langle\pi^-|\bar{d}\gamma^\mu c|D^0\rangle
   =f_+ (q^2)(p_{D^0}+p_{\pi^-})^\mu + f_- (q^2)(p_{D^0}-p_{\pi^-})^\mu,\\
   &q\equiv p_{D^0}-p_{\pi^-}\qquad.
   \end{aligned}
\end{equation}

The imaginary parts of penguin diagram arise from a cut on the internal-line
particles which involves on-shell particles and thus long-distance physics,
so it is difficult to estimate. Nevertheless, we can first calculate the penguin diagram by
assuming that the momentum of gluon is spacelike which is calculable,
then carefully analytically continue
the momentum to timelike to extract the imaginary part. While the result is not
so accurate as in QED, it still can be considered as a reasonable estimation.
The result is
\begin{equation}
\begin{aligned}
  \mathrm{Im}&(\mathcal{M}_P^{\mathrm{SM}}(D^0\to \pi^+ \pi^-))\\
  &=i\alpha_s(\mu)\frac{-2}{27} \frac{G_F}{\sqrt{2}}
   [-\langle\pi^+|\bar{u}\gamma^\mu \gamma_5 d|0\rangle
   \langle\pi^-|\bar{d}\gamma^\mu c|D^0\rangle
   +2\langle\pi^+|\bar{u}\gamma^5 \gamma_5 d|0\rangle
    \langle\pi^-|\bar{d} c|D^0\rangle]\\
  &\approx\alpha_s(\mu)\frac{2}{27}\frac{G_F}{\sqrt{2}}
   m_D^2f_+(m_\pi^2)f_\pi\Big[-1+\frac{2m_\pi^2}{(m_c-m_d)(m_u+m_d)}\Big]~,~
\end{aligned}
\label {eq: penguin}
\end{equation}
where $\mu$ is the typical energy scale in this transition.  By
substituting Eqs. (\ref {eq: parameter1}),
(\ref {eq: tree}) and (\ref {eq: penguin}) into
Eq. (\ref {eq: direct cp}), the final expression is obtained,
\begin{equation}
\begin{aligned}
   A_{CP}^{\mathrm{dir}}(D^0&\to \pi^+ \pi^-)\\
  &=\alpha_s(\mu)\frac 4 {27} \Big[-1+\frac{2m_\pi^2}{(m_c-m_d)(m_u+m_d)}\Big]
   \frac{\mathrm{Im}(V_{ub}V_{cb}^*)}{V_{ud}V_{cd}^*}\\
   &=\alpha_s(\mu)\frac 4 {27}\Big[-1+\frac{2m_\pi^2}{(m_c-m_d)(m_u+m_d)}\Big]
   \frac{A^2\lambda^5\eta}{\lambda(1-\lambda^2/2)}\\
   &\simeq0.0086\%,
  \end{aligned}
\end{equation}
and
\begin{equation}
\begin{aligned}
   A_{CP}^{\mathrm{dir}}(D^0\to &K^+ K^-)\\
  &=\alpha_s(\mu)\frac 4 {27}\Big[-1+\frac{2m_\pi^2}{(m_c-m_s)(m_u+m_s)}\Big]
   \frac{\mathrm{Im}(V_{ub}V_{cb}^*)}{V_{us}V_{cs}^*}\\
   &=\alpha_s(\mu)\frac 4 {27}\Big[-1+\frac{2m_\pi^2}{(m_c-m_s)(m_u+m_s)}\Big]
   \frac{A^2\lambda^5\eta}{\lambda(1-\lambda^2/2)}\\
   &\simeq-0.0087\%,
\end{aligned}
\end{equation}
where we have taken $\mu=m_c$, $\alpha_s(\mu)=\alpha_s(m_c)=0.396$, and
$\lambda=0.2253$, $A=0.808$, $\eta=0.341$, $m_K=493.677 ~\mathrm{MeV}$,
$m_\pi=140 ~\mathrm{MeV}$, $m_s(m_c)=122 ~\mathrm{MeV}$, $m_c=1290 ~\mathrm{MeV}$,
$m_d(m_c)=6.1 ~\mathrm{MeV}$, $m_u(m_c)=3.05 ~\mathrm{MeV}$ \cite{pdg,Buras}.
The U-spin relation
$A_{CP}^{\mathrm{SM}}(D^0\to K^+ K^-)= -A_{CP}^{\mathrm{SM}}(D^0\to \pi^+ \pi^-)$
is guaranteed by the approximated $SU(3)_F$ symmetry.
Finally the difference between  $A_{CP}^{\mathrm{SM}}(D^0\to K^+ K^-)$
and $A_{CP}^{\mathrm{SM}}(D^0\to \pi^+ \pi^-)$ is
\begin{equation}
   \bigtriangleup A_{CP}^{\mathrm{dir}}(\mathrm{SM})=-0.02 \%~.~
\end{equation}

While uncertainties due to nonperturbative QCD might be considerable \cite{Buras,Buras2},
the experimental central value of $\bigtriangleup A_{CP}^{\mathrm{dir}}$
at the LHCb is still difficult to be understood within
the SM. It is well known that CP violation in $D$ meson decays is a clean
way to probe new physics, which has drawn many attentions
\cite{review2,Ddecay,CPRPV,NPCPV,Ddecay2,Dlesson,Dmeson,DDmix}.
In the light of recent experimental result of $ \bigtriangleup A_{CP}^{\mathrm{dir}}$, it is
expected that such kind of new physics would enhance direct CPV in charm
quark decays \cite{3,4,5,6},
while leaving beauty and strange quarks nearly unaffected, it will be
shown that RPV SUSY can provide such an opportunity.

In SUSY, the general trilinear RPV interactions are
\begin{equation}
\label{eq:RPV}
  \mathcal{W}_{{\not R}}\ =\epsilon_{\alpha \beta}
  (\frac{1}{2}\lambda_{ijk} L_i^\alpha L_j^\beta E^c_k
  + \lambda'_{ijk} L_i^\alpha Q_j^\beta D^c_k)
  +\frac{1}{2}\lambda''_{ijk}U_i^c D_j^c D_k^c~,~
\end{equation}
where $\lambda_{ijk}=-\lambda_{jik}$, $\lambda''_{ijk}=-\lambda''_{ikj}$,
and $\lambda'_{ijk}$'s are completely free parameters. Here $L$ and $E^c$
( $Q$, $U^c$ and $D^c$ ) correspond respectively to the lepton
doublet and anti-lepton singlet ( quark doublet and antiquark singlet )
left-handed superfields. Charm
quark nonleptonic decays could be induced by $\lambda', \lambda''$
terms \cite{CPRPV,NPCPV}, the relevant Lagrangian is
\begin{equation}
\begin{aligned}
   {\cal L}\supset
   \lambda '_{ijk}\tilde  l_{iL}\bar d_{kR}u_{jL}
   -\frac{1}{2}\lambda ''_{ijk}(\widetilde{d}_{kR}^* \bar u_{iR} d_{jL}^c
   +\widetilde{d}_{jR}^* \bar u_{iR} d_{kL}^c)+\ \mbox{h.c.}\ .
\end{aligned}
\end{equation}
For simplicity, baryon number conservation would be assumed, specifically only $\lambda'$
terms would be taken into account.  The new charm quark decay diagrams are
shown in Fig.2.  It is found that the following
requirements are essential to understand the LHC-b CPV anomaly:

1) Among various $\lambda '_{ijk}$'s, only two
terms would be introduced,
$\lambda '_{112}$ and $\lambda '_{122}$, while $\lambda '_{112}$ is
real and $\lambda '_{122}$ is complex.

2) Furthermore, the following relation is assumed,
\begin{equation}
   \frac{\mathrm{Im}(\lambda'_{122}\lambda_{112}^{'*})}
   {\widetilde{m}_{e}^2}=\frac{\lambda_{112}'\mathrm{Im}(\lambda'_{122})}
   {\widetilde{m}_{e}^2}
   \simeq 40\times\frac{\mathrm{Im}(V_{ub}V_{cb}^*)}{m_W^2}g_2^2~,~
   \label {eq:assumption}
\end{equation}
where $g_2$ is the weak interaction coupling, and the numerical factor
is inferred from the above SM calculation.

Because of Eq.(\ref{eq:assumption}), there exist an interesting corollary:
the new RPV tree diagrams is negligible compared to the SM tree
diagram,
\begin{equation}
\begin{aligned}
   \mathcal{M}_T^{\mathrm{RPV}}(D^0\to K^+ K^-)
   \sim \frac{\lambda'_{122}\lambda_{112}^{'*}}{\widetilde{m}_{e}^2}
  &\sim 40\times\frac{\mathrm{Im}(V_{ub}V_{cb}^*)}{m_W^2}g_2^2\\
  &\ll \frac{V_{us}V_{cs}^*g_2^2}{m_W^2}
   \sim\mathcal{M}_T^{\mathrm{SM}}(D^0\to  K^+ K^-),
   \end{aligned}
   \label {eq:suppression1}
\end{equation}
and
\begin{equation}
   \mathcal{M}_T^{\mathrm{RPV}}(D^0\to\pi^+ \pi^-)
   \sim \frac{\lambda'_{121}\lambda_{111}^{'*}}{\widetilde{m}_{e}^2}=0,
    \label {eq:suppression2}
\end{equation}
as a result, RPV contributions to the branching ratios of various $D$
and $K$ decays would be negligible compared to their SM decay modes.

Up to now, all necessary ingredients have been prepared. The
calculations are direct. First, consider the $D^0\to\pi^+ \pi^-$
transition, the total amplitude is

\begin{equation}
\begin{aligned}
   &\mathcal{M}(D^0\to\pi^+ \pi^-)
   =\alpha_T^{\mathrm{SM}} \mathcal{M}_T^{\mathrm{SM}}
   +\alpha_T^{\mathrm{RPV}} \mathcal{M}_T^{\mathrm{RPV}}
   +\alpha_P^{\mathrm{SM}} \mathcal{M}_P^{\mathrm{SM}}
   +\alpha_P^{\mathrm{RPV}} \mathcal{M}_P^{\mathrm{RPV}}~, \\
\end{aligned}
\end{equation}
where
\begin{equation}
   \alpha_P^{\mathrm{RPV}} (D^0\to\pi^+ \pi^-)
   =\lambda'_{122}\lambda_{112}^{'*} ~.~
\end{equation}

Because of Eq. (\ref{eq:suppression2}), the total direct CP asymmetry
in $D$ meson can be simplified as
\begin{equation}
   A_{CP}^{\mathrm{dir}}(\mathrm{SM+RPV})\simeq
   \frac{-2\big[
  \mathrm{Im}(\alpha_{P}^{\mathrm{SM}})
   \mathrm{Im}(\mathcal{M}_P^{\mathrm{SM}})
   +\mathrm{Im}(\alpha_{P}^{\mathrm{RPV}})
   \mathrm{Im}(\mathcal{M}_P^{\mathrm{RPV}})
   \big]}
   {\alpha_T^{\mathrm{SM}} \mathcal{M}_T^{\mathrm{SM}}}~.~
\end{equation}

\vspace{1pt}
\begin{figure}[htbp]
\centering
\includegraphics[width=3.5in]{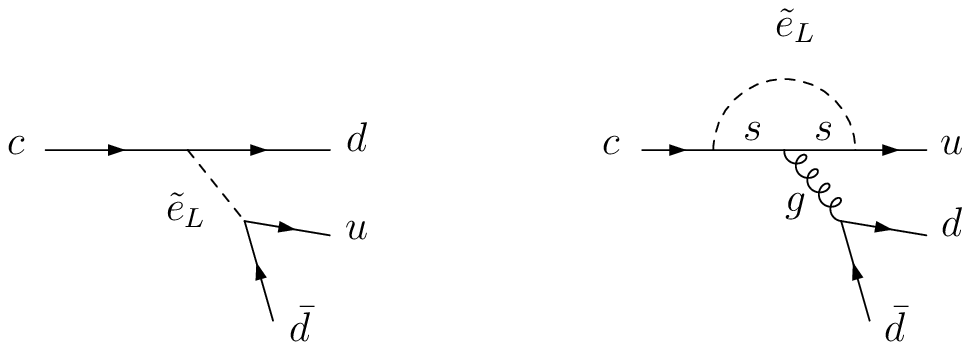}
\caption{$c\rightarrow \bar{d} d u$ tree level and penguin diagrams
in RPV SUSY.}
\end{figure}

Following analogous procedures,
the imaginary part of RPV penguin diagram is
\begin{equation}
\begin{aligned}
   \mathrm{Im}(\alpha_P^{\mathrm{RPV}})&\mathrm{Im}
   (\mathcal{M}_P^{\mathrm{RPV}}(D^0\to \pi^+ \pi^-))\\
   &\approx-\alpha_s(\mu)\frac{m_D^2f_+f_\pi}{108\widetilde{m}_{e_L}^2}
   \Big[-1+\frac{2m_\pi^2}{(m_c-m_d)(m_u+m_d)}\Big]
   \mathrm{Im}(\lambda'_{122}\lambda_{112}^{'*})\\
   &=40\times\alpha_s(\mu)\frac{2}{27}\frac{G_F}{\sqrt{2}}
   m_D^2f_+f_\pi\Big[-1+\frac{2m_\pi^2}{(m_c-m_d)(m_u+m_d)}\Big]\mathrm{Im}(V_{ub}V_{cb}^*)\\
   &=40\times\mathrm{Im}(\alpha_P^{\mathrm{SM}})\mathrm{Im}
   (\mathcal{M}_P^{\mathrm{SM}}(D^0\to \pi^+ \pi^-)).
\end{aligned}
\end{equation}

The total direct CP violation in $D^0\to\pi^+ \pi^-$ transition is now
\begin{equation}
   A_{CP}^{\mathrm{dir}}(D^0\to \pi^+ \pi^-)\simeq 0.35\%.
\end{equation}

Similar calculation results to total CP violation in $D^0\to K^+ K^-$
transition
\begin{equation}
   A_{CP}^{\mathrm{dir}}(D^0\to K^+ K^-)\simeq -0.36\%,
\end{equation}

Now, it is clear that our requirements indeed result in a
considerable enhancement to direct CPV in $D$ decay. In order to be consistent
with current experiments on $K$ mesons and $B$ mesons,
one have to keep new contributions to $K$ and $B$ sectors suppressed.
The $B$ meson decays will not be affected, because only the
$\lambda'_{122}$ and $\lambda'_{112}$ have been introduced.
For $K$ meson, the RPV interactions
$\lambda^{'}_{ijk} \tilde{\nu}_L \bar{d}_R^k d_L^j$ will
generate new diagrams for $s\rightarrow du\bar{u}$ with $s$-quarks being
the internal lines.
However, the direct CPV in $K$ will not be affected, since the internal
quarks can not all be on-shell and hence no imaginary part would arise
through these additional diagrams, hence no extra direct CPV.
In addition, strict experimental constraints in lepton flavor violation
are evaded, since only the first generation of
leptons and their SUSY partners are involved in new interactions.

\section{Conclusions}
In this paper, we investigate supersymmetry without R-parity to
interpret the recent observed large
$\bigtriangleup A_{CP}^{\mathrm{dir}}\equiv A_{CP}(D^0\to K^+K^-)-A_{CP}(D^0\to \pi^+ \pi^-)$
at  LHCb, which corresponds to 3.5 $\sigma$ significance. It is found that a
significant enhancement for the CPV in $D$ meson is feasible after
introducing delicate $R_p$-violation terms $\lambda'_{122}$ and
$\lambda'_{112}$. Phenomennological implications are discussed below:

1) There are many constraints in RPV \cite{RPV}, among them
the following one is of essential relevance to this work,
\begin{equation}
   |\lambda'_{i22}\lambda_{i12}^{'*}|
   < 2.11\times 10^{-5} \Big[\frac{m_{\widetilde{d}_{kR}}}
   {100 \mathrm{GeV}}\Big]^2~.~~
\end{equation}
Combining it with the result shown in Eq. (\ref{eq:assumption}),
we get a relation
\beq
m_{\widetilde{d}_{R}}\geq 13\, m_{\widetilde{e}}.
\eeq
This relation constrains strongly the parameter space of $\rpv-$ SUSY.
After introducing the $\lambda'_{122}$ and $\lambda'_{112}$,
there are some exotic phenomenology \cite{RPV}.
At the LHC, the pair production of the scalar-quark, i.e. process
$pp \rightarrow \tilde{q}\tilde{q}$ and the single production
process $pp \rightarrow \tilde{q} e$
followed by the decay of $\tilde{q}\rightarrow q'+e $ have large cross
section and exotic final states. The reconstructed invariant mass of $\tilde{q}$
from one jet and the electron, and delicate kinematic cuts make
the signal distinguished from the backgrounds which mainly come
from Z+jets \cite{LQPhe}. It is expected that LHC could find the
exotic signal of the $\tilde{q}$ or constrain further the parameter space of the model.

2) For singly Cabibbo suppressed decay modes,
such as $D^+_{s}\to \pi^+ + K^0$, it
is expected that the same order direct CPV will be observed.
Besides the direct CP violation, there is a small enhancement in
the $D^0-\bar{D^0}$ mixing from the new physics.
It is, however, negligible compared to the SM, since the new couplings
are actually CKM suppressed, as shown in Eq.(\ref{eq:suppression1}).
Analogously, the mixing in $K$ system can also be considered as unaffected.

Although it is still far from a complete theoretical description, the
RPV by itself is a very natural way to induce differentiated CP
violations, since $u$-type quarks and $d$-type quarks are treated
differently in RPV terms, which is essential to extend the SM, in which
it is difficult to explain why $D$ meson is more special
than $K$ and $B$ mesons. As the experimental data is accumulating, some more
fundamental mechanisms might be discovered, through which we could
understand why the $\lambda'_{ijk}$'s have taken such specific structures as in
Eq. (\ref{eq:assumption}).

\acknowledgments
We would like to thank Prof. Hai-Yang Cheng and Hua Shao for some helpful discussions.
This work was supported in part by the National Natural
Science Foundation of China under nos. 11075193, 10821504 and 11175251.

\end{document}